\documentclass[12pt]{iopart}

\def\d{{\mathrm d}}

\newcommand{\bm}[1]{\mbox{\boldmath $#1$}}

\def\sign{\mathrm{sign}}
\def\id{\mathrm{1\hspace{-2.8pt}I}}
\def\real{\mathrm{I\hspace{-2.4pt}R}}

\def\Ima{\mathrm{Im}}
\def\Rea{\mathrm{Re}}
\def\lie{\mathcal{L}}

\def\kk{\vec\kappa}
\def\kkf{\bm\kappa}
\def\kkc{\kappa}

\def\comega{\omega}

\def\mm2{\mathcal{W}}

\def\I{\mathrm{\scriptscriptstyle I}}
\def\E{\mathrm{\scriptscriptstyle E}}

\def\spI{(\mm2^\I, g^\I)}
\def\spIB{(\mm2^\I, g^\I,\sup^\I)}
\def\spE{(\mm2^\E, g^\E)}
\def\spEB{(\mm2^\E, g^\E,\sup^\E)}

\def\sup{\Sigma}

\def\curv{\Upsilon}

\def\otho{\Gamma}

\def\tgotho{\tilde\gamma}

\def\stargotho{\star}

\def\axisE{W^\E_2}

\def\ww{W_2}
\def\www{W_3}

\def\ff{\mathcal{F}}
\def\vf{f}
\def\ernst{\mathcal{E}}
\def\ernstf{\Gamma}

\def\jj{{\cal J}}
\def\jjbm{\bm{\mathcal{J}}_{\!\mathrm{\scriptscriptstyle BM}}}
\def\jjj{\underline{\jj}}

\def\RelL2{(\bar\lambda\Lambda+ \lambda\bar\Lambda)}
\def\ImlL2{(\bar\lambda\Lambda- \lambda\bar\Lambda)}
\def\dRelL2{(\bar\lambda\d\Lambda+ \lambda\d\bar\Lambda)}
\def\dImlL2{(\bar\lambda\d\Lambda- \lambda\d\bar\Lambda)}

\def\aa{\mathcal{A}}
\def\bb{\mathcal{B}}

\def\Journal#1#2#3#4#5#6{(#5) ``#6'' {#1} {\bf #2} #3#4}

\def\PRD{\em Phys. Rev. D }

\def\CPAM{\em Comm. Pure Appl. Math.}
 \def\CMP{\em Commun. Math. Phys.}

\def\MPL{\em Mod. Phys. Lett.}

\newtheorem{theorem}{Theorem}
\newtheorem{proposition}{Proposition}

\newtheorem{remarkpro}{Remark}[proposition]

\def\fiproofns{\hfill \rule{2.5mm}{2.5mm}}

\begin{document}

\title[On global models for 
isolated
rotating axisymmetric charged bodies;...]
{On 
global models for 
isolated
rotating axisymmetric charged bodies;
uniqueness of the exterior field}

\author{Ra\"ul Vera}
\address{School of Mathematical Sciences, Dublin City University,
Dublin 9, Ireland}

\begin{abstract}
A relatively recent study by Mars and Senovilla
provided us with a uniqueness result
for the exterior vacuum gravitational field
generated by an isolated distribution of matter
in axial rotation in equilibrium in General Relativity.
The generalisation to exterior electrovacuum
gravitational fields, to include
charged rotating objects, is presented here.
\end{abstract}
\pacs{04.20.Cv, 04.40.Nr, 04.20.Ex}




\section{Introduction}
\label{sec:intro}
A proper understanding of rotating objects in
equilibrium within the context of General Relativity is
fundamental for many astrophysical situations.
Very unfortunately, finding global models for rotating objects
in General Relativity has proven to be extremely difficult, even 
for axially symmetric configurations in equilibrium.
So far, there are no known complete explicit models
for a self-gravitating finite body (with non empty interior)
together with its exterior
except for spherical, and hence non-rotating, configurations.

This work focuses on the theoretical point of view
of the construction of global models for finite objects
by means of the matching of spacetimes:
the whole configuration
is composed of two regions,
one region of spacetime $\spI$ describing the interior (I) of the body and
another $\spE$ describing the exterior field (E), matched across a
hypersurface $\sup$. This matching hypersurface is then a common
boundary of the two regions and corresponds to the limiting surface of
the body at all times.  The two regions can then be treated
independently, taking into account that the two problems will have to
satisfy compatible boundary conditions on $\sup$ imposed by the
matching conditions.
To model the equilibrium state of the rotating configuration, the
whole matched spacetime, and hence each of the interior and exterior regions,
is taken to be strictly stationary.  In addition, it has usually
been assumed that the model is axially symmetric.
To account for the isolation of the body, the exterior region
is required to be asymptotically flat.

In previous works on models of isolated rotating bodies in equilibrium,
only vacuum exteriors have been taken into account.
Nevertheless, compact objects in astrophysics have been also
considered many times as sources of electromagnetic fields
(see \cite{noconv2} and \cite{evsph} and references therein).
In fact, the existence of a net charge, no matter how negligible,
has drastic consequences for the global spacetime structure,
in principle.
The aim of the present work is to revisit the exterior
problem, and generalise the uniqueness results found
for the vacuum case in \cite{MASEuni} to include
stationary and axisymmetric electromagnetic
fields without sources (electrovacuum fields).
The main result 
deals with the uniqueness of the exterior electrovacuum
field generated by an isolated distribution of charged matter
in axial rotation and in equilibrium.
It is worth noticing that 
spherically symmetric global models for 
stellar collapse to include charged radiating stars and voids
has been already considered by several authors during the past three decades
(see \cite{Ori} and \cite{evsph} and references therein).

The exterior 
stationary and axisymmetric Einstein-Maxwell
electrovacuum problem consists of an elliptic PDE
system, known as the Ernst equations,
for a couple of complex functions \cite{sol,heusler};
the electromagnetic potential
$\Lambda$ and the so-called Ernst potential $\ernst$,
which carry all the information about the exterior geometry and
electromagnetic field.
It is known that the matching of a \emph{vacuum} exterior with a given
interior, prescribing the identification of the
interior and the exterior across $\sup$ if required, 
determines: (i) the existence of the matching hypersurface, and if so,
its form as seen from the interior, generally,
(ii) the matching hypersurface as seen from the exterior,
and (iii) the values of $\ernst$ up to an arbitrary
additive imaginary constant, say $i \comega$,
and the normal derivative of $\ernst$ there \cite{MASEuni,conv}.
This data constitutes a
set of boundary conditions of Cauchy type for $\ernst$ on $\sup$
that depend on the parameter $\comega$.

In order to accommodate an electrovacuum exterior,
it is necessary to, first, generalise the spacetime matching
conditions for a vacuum exterior,
also taking into acount the matching conditions for
the electromagnetic field \cite{jackson} (see also \cite{evsph}).
For an object without surface charges\footnote{This assumption
is not really necessary, since one could simply consider a surface
charge and then include that as information that has to be given
with the rest of the interior configuration.}
the electromagnetic field has to be continuous across $\sup$, and hence
the interior quantities determine the electromagnetic potential $\Lambda$
up to an additive complex constant $\lambda$.
Here it will be shown how,
when considering an electrovacuum exterior,
the sets (i) and (ii) remain formally unchanged
whereas the rest of matching conditions
constitute a generalised new set (iii) of Cauchy boundary
data for $\ernst$ 
on $\sup$ in terms of the interior quantities and
involving 
$\comega$ and $\lambda$.
Summing up, given the interior, the matching conditions
determine Cauchy boundary data for $\ernst$ and $\Lambda$ on $\Sigma$
up to three degrees of freedom, given by $\comega$ and $\lambda$.

Due to the elliptic character of the exterior electrovacuum problem,
the Cauchy boundary data,
together with the decaying conditions at infinity,
constitute an overdetermined set of boundary conditions
for the exterior electrovacuum problem.
Nevertheless,
as shown later, the three parameters encoded in $\comega,\lambda$
are in principle physically relevant,  and therefore,
the exterior problem is overdetermined but, a priori, not unique.
The proof of the uniqueness of the exterior solution
has to follow two steps, then. In a first step
it is shown that given the values of $\ernst$ and $\Lambda$ on $\sup$
(Dirichlet boundary data), the exterior field is unique.
On a second step, it is shown that the completion to a whole
set of Cauchy boundary data of a Dirichlet boundary data set that depends on
$\comega,\lambda$ determines the values of $\comega$ and $\lambda$,
and hence the Dirichlet data, provided that the solution exists.

The paper is structured as follows. Section \ref{sec:ext}
is devoted to a brief review concerning the electrovacuum exterior problem.
The dependence on the free parameters
$\comega$ and $\lambda$ of the Cauchy boundary conditions on $\sup$
is determined in Section \ref{sec:bc}.
For the sake of brevity in the main text,
the full determination of the matching conditions is left
to \ref{sec:matching}.
Section \ref{sec:dirichlet} deals then with the proof of the uniqueness
for the Dirichlet problem. Finally, Section \ref{sec:fixing} is devoted
to showing that the full set of Cauchy boundary data determine
the values of $\comega$ and $\lambda$ if the solution exists.
The statement of the final result is left to a conclusions section.


The units all throughout the paper are chosen so that $G=c=1$. 

\section{The exterior problem}
\label{sec:ext}
For completeness and to fix some notation,
this section is devoted to a brief review of the
stationary and axisymmetric electrovacuum problem.
Let me refer to e.g. \cite{sol,heusler} for the details.

Given the electromagnetic field described by the 2-form $\bm F$,
it will be convenient to use the self dual 2-form $\bm\ff$ defined as
\begin{equation*}
\bm\ff\equiv \bm F + i * \bm F,
\end{equation*}
where $*$ stands for the Hodge dual.
The self-dual 2-form $\bm\ff$ is then completely determined by
an arbitrary non-null vector $\kk$ and the projection $\bm \vf_{\kk}
\equiv \bm\ff(\cdot,\kk)$, which clearly
obeys $\bm \vf_{\kk}(\kk)=0$, by
\begin{equation*}
\bm \ff =\frac{1}{N_{\kk}} \left[
\bm \vf_{\kk} \wedge \kkf + i * (\bm \vf_{\kk} \wedge \kkf)\right],
\end{equation*}
where we have defined $ N_{\kk} \equiv \kk\cdot \kk(\neq 0)$,
and $\wedge$ denotes the exterior product
(see e.g.~\cite{sol}).
The real and imaginary parts of the
form $\bm f_{\kk}$ are usually
denoted by $\bm E_{\kk}$
and $\bm B_{\kk}$, and for unit timelike $\kk$, they
correspond to the electric and magnetic
parts of $\bm F$ with respect to $\kk$, respectively,
i.e. \cite{sol}
\begin{equation*}
\bm \vf_{\kk}=\bm E_{\kk} + i \bm B_{\kk}.
\end{equation*}
The Maxwell equations in terms of $\bm\ff$ take the form
\begin{equation*}
\d \bm\ff= 4\pi i * \bm j,
\end{equation*}
where $\bm j$ is the electromagnetic current source,
and the energy-momentum reads
\begin{equation}
T_{\mu\nu}=\frac{1}{8\pi} \ff_{\mu \alpha} \bar\ff_\nu {}^\alpha,
\label{eq:tmunu}
\end{equation}
where the bar denotes the complex conjugate.

Let us consider a strictly stationary\footnote{In global terms, we demand that
the manifold admits an everywhere timelike Killing vector with
complete orbits.}
and axisymmetric 
\cite{sol}
asymptotically flat, causal
and simply connected $C^3$ spacetime with connected boundary
$\spEB$,
containing an 
electrovacuum field $\bm\ff$ invariant under the same group of isometries,
i.e. the stationary and axisymmetric pair $(g^\E,\bm \ff)$ is a
solution of the Einstein-Maxwell equations without sources ($\bm j=0$).
The intrinsically defined axial Killing vector field
\cite{maseaxconf} (see also \cite{sol})
will be denoted by $\vec\eta$, and the axis of symmetry by $\axisE$,
which is non-empty by assumption.
$\vec\xi$ will denote an everywhere timelike
Killing vector field which together with $\vec\eta$ generate
the necessarily Abelian \cite{carter70} (see also \cite{sol})
$G_2$
group of isometries that act on timelike surfaces $T_2$.
It is also well known that the $G_2$ on $T_2$ group
of isometries in a stationary and axisymmetric
electrovacuum spacetime
must act orthogonally transitively (see e.g. \cite{sol}).
These properties,
together with the vanishing of the trace of the energy-momentum
tensor for the electromagnetic field, imply the existence
of a coordinate system $\{t,\phi,\rho,z\}$
with ranges within $\rho> 0,t,z\in \real, 0\leq\phi<2\pi$,
the so-called Weyl canonical coordinates, in which the
line-element for $\mm2^\E\setminus \axisE$ reads locally \cite{sol}
\begin{equation}
  \label{eq:ds2e1}
  \d s^2=-e^{2U}\left(\d t+ A \d\phi\right)^2+e^{-2U}
\left[e^{2k}\left(\d\rho^2+\d z^2\right)+
\rho^2 \d\phi^2\right],
\end{equation}
where $U$, $A$, $k$ are functions of $\rho$ and $z$,
the axial Killing vector
is given by $\vec\eta=\partial_\phi$, the axis
of symmetry $\axisE$ is located at the limit $\rho\rightarrow 0$,
and $\vec\xi=\partial_t$.
Moreover, the coordinate $t$ can be chosen to
measure proper time of an observer at infinity. In that case
the Killing vector $\vec\xi$ is intrinsically defined by being
unit at infinity \cite{sol}.
Within this setting, the coordinate freedom in (\ref{eq:ds2e1}) consists only
on trivial constant shifts of $t,\phi$ and $z$. 

In order to have our $G_2$ group defining complete orbits,
the Killing vectors $\{\vec\xi,\vec\eta\}$ are required to be
tangent to $\sup^\E$. In other words, we demand that $\sup^\E$
preserves the stationarity and axial symmetries
\cite{mps}.
As a consequence, $\sup^\E$ is everywhere timelike by assumption.
Furthermore, since $\mm2^\E$ is assumed simply connected, connectedness
of $\sup^\E$ implies that the slices of constant $t$ of $\sup^\E$ 
are homeomorphic to two-spheres.

The existence of the Weyl canonical
coordinates \emph{globally} on $\mm2^\E\setminus \axisE$
had been ensured in relation with the black hole uniqueness theorems
\cite{carterbh} (see also \cite{weinstein1}),
where $\sup^\E$ lies on the horizon,
so that $\rho|_{\sup^\E}=0$.
In the present case, though, the boundary $\sup^\E$ lies on the surface
of the object, and therefore those results cannot be used in principle.
Nevertheless, the good behaviour of the function $\rho$ can be also
ensured when the boundary $\sup^\E$ at constant $t$
is an axially symmetric surface
homeomorphic to a two-sphere, as it is the case here.
Global Weyl coordinates had been used in previous works
on global models of finite objects, but without proof \cite{MASEuni,conv}.
Due to its length, and not to overwhelm this paper,
it has been preferable to leave the proof on the existence of the
global Weyl coordinates in the present scenario
to a separate work \cite{rho}.
Taking global Weyl coordinates (\ref{eq:ds2e1}) for $\spEB$,
the axially symmetric hypersurface $\sup^\E$
is thus given parametrically by \cite{MASEuni,conv,rho}
\begin{equation}
  \label{eq:mhparam}
  \sup^\E:\{t=\tau,\phi=\varphi,\rho=\rho(\mu ),z=z(\mu )\}.
\end{equation}
%

Denoting by $\kk$ any of the two Killing vectors $\vec\xi$ or $\vec\eta$,
the invariance of the electromagnetic field $\bm F$
with respect to $\kk$, equivalent to
$\lie_{\kk} \bm\ff=0$, implies the existence
of a complex scalar potential $\Lambda_{\kk}$ 
such that (see e.g. \cite{sol,heusler})
\[
\d \Lambda_{\kk}=\bm \vf_{\kk}
\]
for either $\kk$. Moreover, for
stationary and axisymmetric electrovacuum fields,
$\bm \vf_{\kk}$ must be orthogonal to the orbits, and thus
$\Lambda_{\kk}=\Lambda_{\kk}(\rho,z)$ \cite{sol,heusler}.

The Einstein-Maxwell equations free of sources
in the simply connected manifold $\mm2^\E$ imply, in turn, the existence of a
real scalar $\Omega_{\kk}$, the so-called twist potential, such that
\cite{harrison} (see e.g. \cite{sol})
\begin{eqnarray}
  \d \Omega_{\kk}=\bm w_{\kk} - i \left(\Lambda_{\kk} \d \bar\Lambda_{\kk}-
\bar\Lambda_{\kk} \d \Lambda_{\kk}\right),
\label{eq:domega}
\end{eqnarray}
where
\(
\bm w_{\kk}\equiv * (\kkf \wedge \d \kkf)
\)
is the twist 1-form of $\kk$.
Recalling the definition 
\(
N_{\kk}\equiv (\kk \cdot \kk),
\)
the so-called Ernst potential $\ernst_{\kk}$ (with respect to $\kk$) reads
\cite{ernst68b,sol,carterbh}
\[
\ernst_{\kk} = - N_{\kk} - \Lambda_{\kk} \bar\Lambda_{\kk} + i \Omega_{\kk}.
\]
The Einstein-Maxwell equations now reduce to the elliptic system
of equations
for the potentials $\Lambda_{\kk}$ and $\ernst_{\kk}$,
known as the Ernst equations,
plus a quadrature for the remaining function in the metric
($k(\rho,z)$ in (\ref{eq:ds2e1})) in terms of the potentials
(see \ref{sec:matching} for the explicit expressions).
Dropping the $\kk$ subindexes, the Ernst equations read \cite{sol,heusler}
\begin{eqnarray}
  &&N \delta^{ij}\partial_i(\rho \partial_j \ernst)+
\rho\ \delta^{ij}\partial_i\ernst
\left(\partial_j\ernst + 2 \bar\Lambda\partial_j \Lambda\right)=0,
\label{eq:e1}\\
  &&N \delta^{ij}\partial_i(\rho \partial_j \Lambda)+
\rho\ \delta^{ij}\partial_i\Lambda
\left(\partial_j\ernst + 2 \bar\Lambda\partial_j \Lambda\right)=0,
\label{eq:e2}
\end{eqnarray}
where $N=-(\ernst+\bar\ernst+2\Lambda\bar\Lambda)/2$,
indices $i,j$ correspond to $\{\rho,z\}$,
and $\delta_{ij}$ represents the $2\times 2$ identity. 
The solutions of the Ernst equations for $\kk=\vec\xi$ and
$\kk=\vec\eta$, that is $(\ernst_{\vec\xi},\Lambda_{\vec\xi})$
and $(\ernst_{\vec\eta},\Lambda_{\vec\eta})$, are called conjugated
to one another and are bi-uniquely related (see \cite{heusler}).

Given either $(\ernst_{\kk},\Lambda_{\kk})$ solution of the Ernst equations
all the information of the exterior electrovacuum solution 
is recovered.
In particular, the metric for $\spE$ in the form (\ref{eq:ds2e1})
is directly obtained by taking 
$\kk=\vec\xi\, (=\partial_t) $:
$U$ is obtained from
\begin{equation}
  \label{eq:UN}
N_{\vec\xi}=-e^{2U},
\end{equation}
and $A$ is determined, up to a constant, by the quadrature
\begin{equation}
  \label{eq:Aomega}
  \d A=-\frac{\rho}{N^2_{\vec\xi}} \stargotho\bm w_{\vec\xi},
\end{equation}
where $\stargotho$ is used to denote the particular Hodge dual
on the $\{\rho,z\}$ two-plane, so that
$\stargotho\d z=-\d \rho, \stargotho\d \rho=\d z$, and
$\bm w_{\vec\xi}$ is in turn obtained from taking
the imaginary part of $\d \ernst_{\vec\xi}$ and using (\ref{eq:domega}).
The function $k$ is then fixed, up to an additive constant
by the aforementioned quadratures involving
$\ernst_{\vec\xi}$ and $\Lambda_{\vec\xi}$ \cite{sol}.

The boundary conditions for the exterior problem consists
of the conditions on the
boundary associated to the surface of the body
plus decaying conditions 
at infinity.
Asymptotic flatness on $\spE$ determines the behaviour
of the Ernst potential at infinity \cite{walter84}
(see also \cite{bmg,heusler} and references therein) and,
in adition,
requires asymptotic conditions on the electromagnetic field~\cite{walter84}.
The conditions at infinity read
\begin{eqnarray}
  \label{eq:uomegainfinity}
  &&U_{\vec\xi}=-M r^{-1}+O(r^{-2}),~~ \Omega_{\vec\xi}=-2 z Jr^{-3}+O(r^{-3}),\\
  &&\Lambda_{\vec\xi}=  -Q r^{-1}+O(r^{-2}),\label{eq:potinfinity}
\end{eqnarray}
where $r=\sqrt{\rho^2+z^2}$,
for some constants $M$, $J$ and $Q$.

The next section is devoted to the boundary conditions
on the hypersurface
$\sup$ that result from the matching conditions with a given
stationary and axisymmetric interior.

\section{The matching conditions: boundary conditions on $\sup$}
\label{sec:bc}
Let us consider a given stationary and axisymmetric
spacetime with boundary $\spIB$ 
containing an electromagnetic field
$\bm\ff^\I$, and describing the interior region matched
across the stationary and axisymmetric (and thus timelike)
hypersurface $\sup (\equiv \sup^\I=\sup^\E)$ to
the stationary and axisymmetric
asymptotically flat electrovacuum spacetime $\spEB$.
Since we do not consider any superficial charge on the surface
of the body, the usual junction conditions of electrodynamics \cite{jackson}
imply that the electromagnetic field has to be continuous across
the timelike hypersurface $\sup$,
and thus  (see e.g. \cite{evsph})
\[
  \bm \ff^\I|_{\sup}=\bm \ff|_{\sup}.
\]
This clearly leads to a relation for
the gradient of the electromagnetic potential at the exterior
with respect to any vector field $\kk$, $\Lambda_{\kk}$,
and the interior quantities, 
given by
\begin{equation}
\label{eq:dlambda_exp}
\dot\Lambda_{\kk}|_{\sup}={e_3}^\alpha\ff^\I_{\alpha\beta}\kkc^\beta|_{\sup},
~~~
\vec n(\Lambda_{\kk})|_{\sup}=n^\alpha\ff^\I_{\alpha\beta}\kkc^\beta|_{\sup},
\end{equation}
where the dot denotes differentiation with respect to
$\mu$, $\vec e_3$ denotes the corresponding vector field,
tangent to $\sup$, and $\vec n$ is normal (not necessarily unit)
to $\sup$. 
It must be stressed that the absence of a surface charge
is made here only for simplicity reasons. One can always consider
a surface charge and simply add it to the information required from the
interior region. In any case, the data $\bm \ff|_{\sup}$ would
then be fixed by the electromagnetic matching conditions (see \cite{evsph}).
In short, given the interior electromagnetic field, the matching fixes
\begin{equation}
  \label{eq:mcem}
  \d \Lambda_{\kk}|_{\sup},
\end{equation}
for either $\kk$ for the exterior problem.
As a consequence, $\Lambda_{\kk}|_{\sup}$
are fixed up to
transformations of the form
 \begin{equation}
   \label{eq:mclambda}
   \Lambda_{\kk}|_{\sup}\rightarrow
   \Lambda'_{\kk}|_{\sup}=\Lambda_{\kk}|_{\sup}+\lambda_{\kk},
 \end{equation}
where $\lambda_{\kk}$ is an arbitrary complex constant
for each respective $\kk$.

Regarding the matching of the spacetime $\spE$ with a given $\spI$,
it is proven in \ref{sec:matching}
(see also \cite{MASEuni,conv} for particular cases) that the matching
conditions determine the existence and thereby the form of
the matching hypersurface
$\sup\equiv\sup^\I=\sup^\E$
as seen from both sides, in the general case,
together with the values of the metric
functions $U$ and $A$ (see (\ref{eq:ds2e1})) 
and their normal derivatives on $\sup$.
It is worth noticing that the
additive constants in $A$ and $k$ (see above) are then determined.
Also, one still has to keep in mind that
it has been assumed that the identification
of the exterior and interior across $\sup$ has been prescribed,
in order to fix two extra degrees of freedom introduced
by the matching procedure, that correspond to
the identification on $\sup$ of the Killing vector $\partial_t$ with
a linear combination of two Killings vectors in the interior,
see \ref{sec:matching}
and \cite{MASEuni,mps,conv}.
The data on $\sup$ 
for the exterior problem
introduced by the matching with a given interior is thus given by
\begin{eqnarray}
   \label{eq:mc1}
   U|_{\sup},~~&  \d U|_{\sup},\\
   A|_{\sup},&  \d A|_{\sup},
 \label{eq:mc2}
\end{eqnarray}
which constitute Cauchy boundary data for the exterior problem.

It is now straightforward to show how
the data given by (\ref{eq:mc1})-(\ref{eq:mc2}) and (\ref{eq:mclambda})
translates
onto data for $\ernst$.
To begin with, the data (\ref{eq:mc1})-(\ref{eq:mc2}) implies that,
given the interior, the matching conditions fix
\begin{equation}
  \label{eq:mcN}
  N_{\kk}|_{\sup},~~
  \d N_{\kk}|_{\sup}
\end{equation}
for either $\kk$.
Let us now consider the case $\kk=\vec\xi$.
Using the data (\ref{eq:mc2}) and (\ref{eq:mcN}), the relation
(\ref{eq:Aomega}) implies that the interior geometry fixes
$\bm w_{\vec\xi}|_\sup$. Therefore, from (\ref{eq:domega}),
and taking into account the freedom generated by $\lambda$
as introduced in (\ref{eq:mclambda}),
the matching conditions fix $\d \Omega_{\vec\xi}|_\sup$
up to transformations of the form
 \begin{equation}
   \label{eq:mcomega1}
   \d \Omega_{\vec\xi}|_{\sup}\rightarrow
   \d\Omega'_{\vec\xi}|_\sup=
   \d \Omega_{\vec\xi}|_{\sup}-i
\left(\lambda_{\vec\xi}\,\d \bar\Lambda_{\vec\xi}-\bar\lambda_{\vec\xi}\,
  \d \Lambda_{\vec\xi} \right)|_\sup,
\end{equation}
and thus fix $\Omega_{\vec\xi}|_\sup$ up to
\begin{equation}
   \label{eq:mcomega2}
\Omega_{\vec\xi}|_{\sup}\rightarrow
\Omega'_{\vec\xi}|_\sup=
   \Omega_{\vec\xi}|_{\sup}+\comega_{\vec\xi} - i
(\lambda_{\vec\xi} \bar\Lambda_{\vec\xi}-   \bar\lambda_{\vec\xi} \Lambda_{\vec\xi})|_\sup, 
 \end{equation}
where $\comega_{\vec\xi}$ is a real arbitrary constant.
Although for the sake of clarity these relations
have been explicitly deduced here for $\kk=\vec\xi$,
it can be shown that the matching conditions fix, in fact,
\begin{equation}
  \label{eq:mctwist}
  \bm w_{\kk}|_{\sup} 
\end{equation}
for both $\kk$, and thus (\ref{eq:mcomega1})-(\ref{eq:mcomega2})
hold for both $\kk$.
Combining all the above,
the Cauchy data on $\sup$ for the Ernst potential is fixed
up to transformations of the form (dropping the $\kk$ subindexes)
\begin{eqnarray}
\label{eq:mcernst}
\ernst|_{\sup}&\rightarrow&
\ernst'|_\sup=
   \ernst|_{\sup}-\lambda\bar\lambda+i\comega
-2\bar\lambda \Lambda|_{\sup},
 ~\mbox{ and }\\
\label{eq:mcdernst}
\d\ernst|_{\sup}&\rightarrow&
\d\ernst'|_{\sup}=
   \d\ernst|_{\sup}-2\bar\lambda\d \Lambda|_\sup.
 \end{eqnarray}

Due to the elliptic character of the Ernst equations system,
the Cauchy data induced by the matching conditions
overdetermine the problem.
Nevertheless, as shown above, there are still three degrees of freedom
on that data, generated by $\lambda$ and $\comega$. 
As remarked in \cite{MASEuni}, the parameter $\comega$ is relevant,
in principle, for the exterior problem,
because the freedom in $\Omega$ has been already used to set its
asymptotic behaviour (\ref{eq:uomegainfinity}).
For the same reason, expression (\ref{eq:mcomega2}) makes clear
that the parameter $\lambda$ is also relevant.
All this means that although the exterior problem for a given
interior is an overdetermined problem, the problem
is still not unique. 
Therefore, in order to solve the uniqueness problem we
have to address the following two questions:
\begin{enumerate}
\item[(a)] uniqueness of the exterior solution $(\ernst,\Lambda)$
given Dirichlet data on $\sup$, i.e. $\{\ernst|_\sup,\Lambda|_\sup\}$,
and
\item[(b)] if a solution $(\ernst,\Lambda)$ for given data
$\{\ernst|_\sup,\Lambda|_\sup,\d\ernst|_\sup,\d\Lambda|_\sup\}
(\lambda,\comega)$, fixed up to the transformations
(\ref{eq:mcernst})-(\ref{eq:mcdernst}), 
exists, do  $\lambda$ and $\comega$ get determined?
In other words, does the Cauchy data determine
$\lambda$ and $\comega$, thereby fixing the Dirichlet data,
if the solution exists?
\end{enumerate}
Clearly, an affirmative answer to both questions
proves the uniqueness of the electrovacuum exterior field given a
charged finite source.

Regarding question (a), the procedure commonly used for
proving uniqueness of Dirichlet problems is based on interpreting
the Ernst equations as the Euler-Lagrange equations for a certain
harmonic map. The proofs rely then, in particular, on
the positivity of the Dirichlet functional associated
to the harmonic map corresponding to the Ernst potential.
This positivity is ensured for both $\kk$ in the vacuum case,
and the proof presented in \cite{MASEuni} regarding question (a)
is based on choosing $\kk=\vec\xi$.
However, in the presence of Maxwell fields (see below),
one needs choosing $\kk=\vec\eta$.
Adapting the proofs of electrovacuum black hole uniqueness theorems
\cite{carterbh,heusler} to the present setting,
Section \ref{sec:dirichlet} is devoted to solve the uniqueness of the
present Dirichlet problem using the Mazur identity approach.
The choice of the Mazur identity approach over the Bunting
identity approach used in \cite{MASEuni} has been made for
convenience, because the Mazur construction presented
in Section \ref{sec:dirichlet} 
is useful in dealing with question (b) in Section \ref{sec:fixing}.


The affirmative answer to question (b) acounts to showing
that if a solution $(\ernst,\Lambda)$ for given data
$\{\ernst|_\sup,\Lambda|_\sup,\d\ernst|_\sup,\d\Lambda|_\sup\}$
exists, then another solution $(\ernst',\Lambda')$
with $\{\ernst'|_\sup,\Lambda'|_\sup,\d\ernst'|_\sup,\d\Lambda'|_\sup\}$
given by $\d\Lambda'|_\sup=\d\Lambda|_\sup$,
(\ref{eq:mclambda}), (\ref{eq:mcernst})-(\ref{eq:mcdernst})
in terms of $\lambda$ and $\comega$ exists only if $\lambda=0=\comega$.
This is done in Section \ref{sec:fixing}.
Since this proof does not need any preference of vector $\kk$,
I have preferred to present the calculations by
choosing $\kk=\vec\xi$, so that the corresponding quantities
that appear have a usual and direct physical meaning.

\section{Uniqueness given Dirichlet data on $\sup$}
\label{sec:dirichlet}
The proof 
follows 
those used in the uniqueness theorems
of black holes (see \cite{carterbh,heusler}), making use of
the very rich intrinsic structure of the Ernst equations,
the only difference being that the horizon is replaced
by the matching hypersurface $\sup$.
In what follows, 
the $\kk$ indices will be omitted for simplicity in the expressions
that hold for both $\kk$, whenever this does not lead to confussion.

Equations (\ref{eq:e1})-(\ref{eq:e2}) can be interpreted as
the Euler-Lagrange equations for the action \cite{heusler}
\[
{\cal S}=4\int_{\mm2^\E}\left(\frac{1}{4N^2}
|\d\ernst+2\bar\Lambda\d\Lambda|^2+\frac{1}{N}|\d\Lambda|^2\right)
\bm\eta,
\]
where $|\bm \theta|^2\equiv g^{\alpha\beta}\theta_\alpha\bar\theta_\beta$
for any 1-form $\bm\theta$,
and $\bm\eta$ is the volume element. 
Taking $\Phi$ to be a hermitian matrix in $SU(2,1)$ defined
by \cite{heusler} ($a,b:1,2,3$)
\begin{equation}
  \label{eq:phi}
\Phi_{ab}\equiv \eta_{ab}+2\ \sign(N)\bar v_a v_b
 \end{equation}
where
$
v_a=(2\sqrt{|N|})^{-1}(\ernst-1,2\Lambda,\ernst+1)
$
and $\eta_{ab}
=\mbox{diag}(-1,1,1)$,
one can define a su$(2,1)$-valued 1-form by\footnote{
Usual matrix
product is denoted by
$(A\cdot B)_{ab}=\sum_c A_{ac} B_{cb}$
and the trace reads
$\Tr A= \sum_a A_{aa}$.
}
\begin{equation}
  \label{eq:jj}
\bm\jj\equiv \Phi^{-1}\cdot \d \Phi.
\end{equation}
The information carried by the
solutions $(\ernst,\Lambda)$ has been now translated onto
$\Phi$. In terms of $\bm \jj$, the above action is rewritten as
\cite{carterbh,smodels,heusler}
\[
{\cal S}=\int_{\mm2^\E}\frac{1}{2}
g_{\alpha\beta}\Tr\left(\jj^\alpha\cdot\jj^\beta\right)
\bm\eta,
\]
for which the variational equation reads \cite{carterbh,smodels,heusler}
\begin{equation}
  \label{eq:jjdiv}
  \nabla^\alpha\jj_\alpha=0.
\end{equation}
Note that the way $\bm \jj$ is defined is not unique.
In fact, although
the definition given above corresponds to that in
\cite{carterbh} after a trivial interchange $\jj_2\leftrightarrow \jj_3$
(see also \cite{heusler}),
later it will be more useful to refer to the construction
presented in \cite{smodels}.

Since the 
object $\Phi$ is stationary and axisymmetric by definition,
the 1-form $\bm \jj$ depends only on $\rho$ and $z$
and furthermore, it 
is ``tangent'' to the surfaces orthogonal to the orbits $T_2$,
i.e. $\jj_\alpha=(0,0,\jj_i)$.
Both $\Phi$ and $\bm \jj$ can therefore
be taken as defined on the Riemannian space $\otho$
that corresponds to the surfaces orthogonal to the orbits
in $\spE$.
Recalling that $\sup$ must be spatially homeomorphic to
a two-sphere, the boundary of the Riemannian
domain $\otho$ of the solutions for $\Phi$  
is given by (see \cite{rho}):
\begin{enumerate}
\item[(i)] A connected curve $\curv_\sup$, that
corresponds to the projection of $\sup$ onto $\otho$,
with ends at the axis,
\item[(ii)] the two segments
that correspond to either (connected) part of the axis, say
$\curv_+$ and $\curv_-$, where $\rho=0$,
\end{enumerate}
so that
$\partial \otho=\curv_\sup\cup \curv_-\cup \curv_+$ is connected.
We will also use $\curv_\infty$ to denote infinity, at
$r^2\equiv\rho^2+z^2\rightarrow\infty$, this is,
the ideal point of $\otho$.
Taking the metric 
$\gamma\equiv |N| g^\E_{ij} \d x^i \otimes \d x^j$
on $\otho$ one has
$\nabla^\alpha \jj_\alpha=|N|\rho^{-1}{}^{(\gamma)}\nabla^i(\rho\jj_i)$,
where ${}^{(\gamma)}\nabla$ denotes the covariant derivative
with respect to $\gamma$, and thus
equation (\ref{eq:jjdiv}) reads 
\begin{equation}
  \label{eq:jjdivg}
    {}^{(\gamma)}\nabla^i(\rho\jj_i)=0
\end{equation}
on $(\otho,\gamma)$.
The Dirichlet boundary data on $\curv_\sup$, i.e. $\Phi|_{\curv_\sup}$,
is obtained by construction from the data
$\{\ernst|_{\curv_\sup},\Lambda|_{\curv_\sup}\}$
which naturally corresponds to
$\{\ernst|_{\sup},\Lambda|_{\sup}\}$.

The first key property for the proofs of the uniqueness theorems
is the positivity of $\Phi$ (and thus, of the action),
which is ensured by taking
$\kk=\vec\eta$
(since $\sign(N_{\vec\eta})=1$ outside the axis)
see \cite{carterbh,heusler}.
All the quantities in the remaining of
this section will be associated to $\kk=\vec\eta$.
Let us define now $X\equiv N_{\vec\eta}$
(which in Weyl coordinates reads
$e^{-2U}\rho^2-e^{2U}A^2> 0$ outside the axis),
$Y\equiv \Omega_{\vec\eta}$
and $A_{\vec\eta}\equiv -Ae^{2U} X^{-1}$. The line-element
(\ref{eq:ds2e1}) can be cast as
\[
\d s^2=-X^{-1}\rho^2 \d t^2+ X(\d \phi+A_{\vec\eta}\d t)^2+X^{-1}e^{2k}
(\d \rho^2+\d z^2).
\]
The problem with this choice is that $X$, 
which appears
explicitly in the action, vanishes
on the axis (located at $\rho=0$),
and therefore a careful analysis there is needed
(see \cite{carterbh}).
It is convenient to change the coordinates
$\{\rho,z\}$ to
prolate spheroidal coordinates $\{x,y\}$, $x>1,|y|< 1$
by $\rho^2=\nu^2(x^2-1)(1-y^2),
z=\nu x y$, where $\nu$ is an arbitrary positive constant,
so that $\d\rho^2+\d z^2=\nu^2(x^2-y^2)
\d s^2_{\tgotho}$
where
$$\d s^2_{\tgotho}=\frac{1}{x^2-1}~\d x^2+\frac{1}{1-y^2}~\d y^2\equiv
\tgotho_{ij}\d \tilde x^i\d \tilde x^j.$$
Taking the metric $\tgotho=e^{-2k}\nu^{-2}(x^2-y^2)^{-2}\,\gamma$ on $\otho$,
due to the conformal invariance
of the equation for $\bm\jj$ (\ref{eq:jjdivg}),
in $(\otho,\tgotho)$ one has
\[\widetilde\nabla^i(\rho \jj_i)=0,\]
where ``tilded'' quantities will refer to the metric $\tgotho$. 
In the $\{x,y\}$ coordinates the boundary $\partial \otho$
is given by
$\curv_\pm:\{y=\pm 1\}$, while
$\curv_\sup$ is a connected curve for
which $x>1$ necessarily that joins $y=1$ with $y=-1$,
and one has $\curv_\infty:\{x\rightarrow\infty\}$.

The core of the proof consists of
considering two different sets of solutions $\Phi_{(1)}$ and
$\Phi_{(2)}$, with corresponding pairs $\bm\jj_{(1)},\bm\jj_{(2)}$
and field variables
$X_{(1)},X_{(2)}$, etc, with common Dirichlet 
data 
on $\curv_\sup$, i.e.
\begin{equation}
\label{eq:dirichlet}
\Phi_{(1)}|_{\curv_\sup}=\Phi_{(2)}|_{\curv_\sup}.
\end{equation}
Defining the differences $\underline f\equiv f_{(1)}-f_{(2)}$,
for any functional $f$ of the variables in $\Phi$,
and \[\Psi\equiv \Phi_{(1)}\cdot \Phi_{(2)}{}^{-1}-\id,\]
where $\id$ is the identity,
so that $\Psi=0\Leftrightarrow \Phi_{(1)}=\Phi_{(2)}$,
the Mazur identity follows (see \cite{carterbh,heusler}):
\[
\widetilde\nabla^i(\rho \Tr \Psi_{,i})
=\rho~ \tgotho_{ij}\Tr\{\jjj^{\dagger i}\cdot\Phi_{(2)}
\cdot \jjj^j\cdot \Phi_{(1)}\}\geq 0,
\]
where the positive semi-definiteness
comes from the fact that $\rho\geq 0$ and $\tgotho_{ij}$ and $\Phi$
are positive definite.
Denoting by $\d\curv^i$ the vector surface element on $\partial\otho$
pointing from $\otho$ outwards,
the Stokes theorem applied to the above identity leads to
\begin{equation}
\label{eq:integral}
\int_{\partial \otho}\rho \Tr\Psi_{,i}\d \curv^i=
\int_{\otho}\rho~ \tgotho_{ij}\Tr\{\jjj^{\dagger i}\cdot\Phi_{(2)}
\cdot \jjj^j\cdot \Phi_{(1)}\}\tilde{\bm\eta},
\end{equation}
as long as $\rho \Tr\Psi_{;i}\d \curv^i=0$
at $\curv_\infty$, and the integrand on the left
is continuous up to $\partial \otho$.

Now, defining $\widehat f\equiv f_{(1)}+f_{(2)}$,
one has \cite{carterbh,carterww,heusler}
\begin{equation}
  \label{eq:trace}
  \Tr\Psi=\frac{1}{X_{(1)}X_{(2)}}
  \left[\underline X^2+2 \widehat X |\underline\Lambda|^2+
    |\underline\Lambda|^4
  +\left[\underline Y+
    \Ima(\underline \Lambda\bar{\widehat\Lambda})\right]^2\right].
\end{equation}
Taking the limits of the values of $X$, $Y$ and $\Lambda$ on the axis
and 
at infinity --with decays given by
(\ref{eq:uomegainfinity})-(\ref{eq:potinfinity}),
but independently of the values of $M$, $J$, and $Q$--
of the configuration
as were computed by Carter \cite{carterww}
(see also \cite{carterbh,heusler}),
one first obtains that $\Tr \Psi$ and $\Tr\Psi_{,i}\d \curv^i|_{\curv_{\pm}}$
are regular ($C^1$) on the axis \cite{carterww}, so that
$\rho\Tr\Psi_{,i}\d \curv^i|_{\curv_{\pm}}=0$, and secondly, 
that
$\rho~ \Tr\Psi_{,i}\d \curv^i|_{\curv_{\infty}}$ also vanishes \cite{carterww}.
As a result, (\ref{eq:integral}) holds and the only
contribution to the integral on the left hand side can come from
$\curv_\sup$.
On the other hand, since the numerator
of $\Tr\Psi$ is quadratic in the differences of the
variables, $\underline X$, etc,
and $\underline \Phi|_{\curv_\sup}=0$ by assumption, one infers
$\d \Tr \Psi|_{\curv_\sup}=0$, and hence
$\rho~ \Tr \Psi_{,i}\d \curv^i|_{\partial\otho}=0$.

Therefore, 
due to the positive semi-definiteness of the integrand
on the right hand side, (\ref{eq:integral}) implies
that $\bm\jjj=0$ in $\otho$ (and thus in $\mm2^\E$).
Since
$\d\Psi=\Phi_{(1)}\cdot \bm \jjj\cdot \Phi_{(2)}{}^{-1}$
by construction,
$\Psi$ is constant all over $\otho$, and thus $\Psi=0$ because
$\Psi|_{\curv_{\sup}}=0$
by assumption (\ref{eq:dirichlet}).
This ends the proof showing that given Dirichlet data
on $\curv_\sup$, and correspondingly on $\sup$,
i.e. $\{\ernst_{\vec\eta}|_{\sup},\Lambda_{\vec\eta}|_\sup\}$,
the solution $(\ernst_{\vec\eta},\Lambda_{\vec\eta})$ of the Ernst
equations in the exterior region $\spE$ is unique.
The correspondence between conjugate solutions
$(\ernst_{\vec\eta},\Lambda_{\vec\eta})$ and $(\ernst_{\vec\xi},\Lambda_{\vec\xi})$
(see~\cite{heusler}) leads to the same result for $\kk=\vec\xi$. 
The main result found in this section can be stated as follows:
\begin{proposition}
\label{res:dirichlet}
Let $(\mm2^\E,g^\E,\bm\ff,\sup)$ 
define a 
simply connected
stationary and axially symmetric electrovacuum spacetime with a connected
boundary preserving the symmetries.
If $\mm2^\E$ is asymptotically flat and the problem is given
Dirichlet boundary data on $\sup$ for the Ernst and electromagnetic
potentials, i.e. $\{\ernst|_{\sup},\Lambda|_{\sup}\}$,
then the solution $(\ernst,\Lambda)$ 
is unique.\fiproofns
\end{proposition}

\section{Fixing the Dirichlet data}
\label{sec:fixing}
The purpose now
is to show how the full set of Cauchy data, i.e. taking into account
$\{\d\ernst|_\sup,\d\Lambda|_\sup\}(\comega,\lambda)$ on top
of the family of
Dirichlet data $\{\ernst|_{\sup},\Lambda|_{\sup}\}(\comega,\lambda)$,
fixes the values of $\comega,\lambda$,
provided that the solution exists. 

The proof presented here makes use of the
divergence free fields $\bm\jj_{\kk}$ (\ref{eq:jj}) choosing $\kk=\vec\xi$.
Decomposing $\bm\jj_{\vec\xi}$ on a basis of su$(2,1)$ one can obtain
eight conserved real 1-forms. 
For convenience, 
the basis used here corresponds to
that presented in \cite{smodels},
so that the eight forms will be presented
as a set of four real 1-forms and two complex 1-forms.
The representation of an element
$Z\in \mbox{su}(2,1)=
\epsilon e+
\delta d +\lambda s +\pi p+
\alpha h+\beta a  $,
where four elements of the basis of the algebra are denoted as
$e,d,s,p$ and the remaining four are encoded
in $h$ and $a$ by ``complexifying'' them,
is chosen as \cite{smodels}
\begin{equation}
\varrho(Z) = \left(
  \begin{array}[c]{ccc}
    \frac{1}{2}\delta-\frac{1}{3}\lambda &-\sqrt{2}\,\bar \alpha & \epsilon\\
    i\sqrt{2}\,\beta & \frac{2}{3}\lambda & i\sqrt{2}\,\alpha\\
    \pi & \sqrt{2}\,\bar\beta & -\frac{1}{2} \delta - \frac{1}{3} \lambda
  \end{array}
\right),
\label{eq:repre}
\end{equation}
where $\alpha$ and $\beta$ are complex contants
and the rest are real.
Nevertheless, one cannot yet revert to the decomposition of the divergence-free
currents in \cite{smodels}, because
the construction of the currents satisfying
(\ref{eq:jjdiv}) is performed in \cite{smodels} in a different way.
In fact, whereas the vacuum case lies on su$(1,1)$ when using
(\ref{eq:phi})-(\ref{eq:jj}), the construction presented in
\cite{smodels} leads to an isomorphic sl$(2,\real)$
in the vacuum case.
Denoting by $\jjbm$ the matrix representation of
the divergence-free currents 
as constructed in \cite{smodels} for $\kk=\vec\xi$, and defining
\[
\sigma\equiv\left(
  \begin{array}[c]{ccc}
    1 & 0 & -i\\
    0 & i\sqrt{2} & 0\\
    1 & 0 & i\\
  \end{array}
\right),
\]
the isomorphism between the two constructions
can be found to be given explicitly by
\begin{equation}
  \label{eq:rel}
  \jjbm=\frac{1}{4}~ \sigma^\dag \cdot \bm \jj_{\vec\xi}\cdot \sigma.
\end{equation}
Using (\ref{eq:repre}) $\jjbm$ is decomposed (see also \cite{smodels} pp.798)
in terms of a basis of four real 1-forms
$\bm\jj^{e},\bm\jj^d,\bm\jj^s,\bm\jj^p$ plus two complex
ones, $\bm\jj^h$ and $\bm\jj^a$,
which corresponds to the elements of the basis
of the algebra su$(2,1)$ listed above, as
\[
\jjbm=\frac{1}{2}\bm\jj^e e+\bm\jj^d d+\bm\jj^s s+\frac{1}{2}\bm\jj^p p
+\frac{1}{2}i\bm\jj^h h+\frac{1}{2}\bm\jj^a a.
\]
The explicit expressions, which can be obtained from the
above construction (\ref{eq:phi})-(\ref{eq:jj})
by making use of (\ref{eq:rel}), are given in \ref{app:jj}.
Using these 1-forms, one defines on $(\otho,\gamma)$
a 
(complex) 1-form 
depending on a $(\ernst,\Lambda)$
configuration plus eight real parameters
as
\begin{eqnarray*}
&&
\bm W(\ernst,\Lambda;c_e,c_d,c_s,c_p,c_{h_1},c_{h_2},
c_{a_1},c_{a_2})
\equiv
\\
&&~~~~
c_e\bm\jj^e+c_d\bm\jj^d+c_s\bm\jj^s+c_p\bm\jj^p+
c_{h_1}\bm\jj^h+ c_{h_2}\bar{\bm\jj}{}^h+
c_{a_1}\bm\jj^a+ c_{a_2}\bar{\bm\jj}{}^a,
\end{eqnarray*}
where clearly all $\bm\jj=\bm\jj(\ernst,\Lambda)$,
which by construction and (\ref{eq:jjdivg})
satisfies
\begin{equation}
\label{eq:divW}
{}^{(\gamma)}\nabla^i(\rho\, W_i)=0.
\end{equation}

The only non-vanishing surface integrals at spatial
infinity 
$\curv_{\infty}$ of the above 1-forms are given by (see e.g. \cite{smodels})
\begin{eqnarray*}
  \int_{\curv_\infty}\rho\,{\jj^d}_i\, \d \curv^i=4 M,~~~~
  \int_{\curv_\infty}\rho\,{\jj^h}_i\,\d \curv^i=
  \int_{\curv_\infty}\rho\,{\jj^a}_i\, \d \curv^i=2 Q,
\end{eqnarray*}
where $M(\ernst,\Lambda)$ and $Q(\ernst,\Lambda)$
relative to a given solution $(\ernst,\Lambda)$
correspond to the mass and the electric charge
of the configuration,
respectively, so that at spacelike infinity one has
(\ref{eq:uomegainfinity})-(\ref{eq:potinfinity}).
As a consequence of (\ref{eq:divW}) we have
\begin{equation}
\fl-\int_{\curv_{\sup}}\rho\,W_i\, \d \curv^i=
\int_{\curv_\infty}\rho\,W_i \d \curv^i=4 M c_d
+2 Q \left(c_{h_1}+c_{a_1}+ c_{h_2}+c_{a_2}\right).
\label{eq:int}
\end{equation}
Now, let us assume two exterior solutions for the same
interior exist $(\ernst_{(1)},\Lambda_{(1)})$,
$(\ernst_{(2)},\Lambda_{(2)})$, with corresponding
$M_{(1)}$, $Q_{(1)}$ and $M_{(2)}$, $Q_{(2)}$ respectively,
such that their Cauchy
boundary data differs by $\lambda$ and $\comega$
as given by the relations (\ref{eq:mcem}), (\ref{eq:mclambda}),
(\ref{eq:mcernst}), and (\ref{eq:mcdernst}): 
\begin{eqnarray*}
&\fl\Lambda_{(2)}|_{\curv_\sup}=\Lambda_{(1)}|_{\curv_\sup}+\lambda,~~~~~~~
&\ernst_{(2)}|_{\curv_\sup}=\ernst_{(1)}|_{\curv_\sup}-\lambda\bar\lambda+
i \comega -2\bar\lambda \Lambda|_{\curv_\sup},\\
&\fl\d \Lambda_{(2)}|_{\curv_\sup}=\d \Lambda_{(1)}|_{\curv_\sup},~~~~
&\d\ernst_{(2)}|_{\curv_\sup}=\d\ernst_{(1)}|_{\curv_\sup}-
2\bar\lambda \d\Lambda|_{\curv_\sup}.\\
\end{eqnarray*}
From these relations, the following equality holds on ${\curv_\sup}$,
\begin{equation}
\bm W(\ernst_{(1)},\Lambda_{(1)};c_e,\ldots,c_s)|_{\curv_\sup}=
\bm W(\ernst_{(2)},\Lambda_{(2)};\hat c_e,\ldots,\hat c_s)|_{\curv_\sup},
\label{eq:ws}
\end{equation}
for a set of eight certain relations
\begin{eqnarray}
\label{eq:relations}
  &&\hat c_e=\hat c_e(c_e,\ldots, c_s; \comega,\lambda),~~~
\ldots,~~~\hat c_s=\hat c_s(c_e,\ldots, c_s; \comega,\lambda),
\end{eqnarray}
which are given explicitly in \ref{app:jj}.
The integration of (\ref{eq:ws}) over ${\curv_\sup}$ using (\ref{eq:int})
leads to
\begin{eqnarray}
\label{eq:final}
  &&4 M_{(1)} c_d
+2 Q_{(1)} \left(c_{h_1}+c_{a_1}+c_{h_2}+c_{a_2}\right)
\nonumber\\
&&~~~~ = 4 M_{(2)} \hat c_d
+2 Q_{(2)} \left(\hat c_{h_1}+\hat c_{a_1}+
\hat c_{h_2}+\hat c_{a_2}\right),
\end{eqnarray}
which has to hold for arbitrary choices of the eight parameters
$\{c_e,\ldots,c_s\}$.
A straightforward calculation, see \ref{app:jj}, leads to the fact that
\[
M_{(1)}+M_{(2)}\neq 0 \Rightarrow \lambda=\comega=0,
\]
(so that in fact $M_{(1)}=M_{(2)}$, $Q_{(1)}=Q_{(2)}$ 
necessarily). It must be stressed here that the case $M_{(1)}+M_{(2)}=0$
does not lead to the same result in general, see \ref{app:jj}.
Nevertheless, and since the mass for physically motivated solutions is
positive, this result basically implies uniqueness of the Cauchy data
for which the exterior field exists: the freedom
in the Cauchy and Dirichlet data generated
by $\lambda$ and $\comega$ is fixed.
To be more precise, we have proven the following:
\begin{proposition}
\label{res:fix}
Let us consider $(\mm2^\E,g^\E,\bm\ff,\sup)$
as in Proposition \ref{res:dirichlet},
with a family of Dirichlet boundary data depending on
three real parameters, i.e.
$\{\ernst|_{\sup},\Lambda|_{\sup}\}(\comega,\lambda)$,
so that it is determined up to transformations of the form
(\ref{eq:mclambda}) and (\ref{eq:mcernst}).
Consider now the completion to Cauchy boundary data by adding
the data $\{\d\ernst|_{\sup},\d\Lambda|_{\sup}\}(\comega,\lambda)$
determined up to transformations of the form $\d\Lambda'|_\sup=\d\Lambda|_\sup$
and (\ref{eq:mcdernst}). If within the $(\comega,\lambda)$-family of
Cauchy data sets there is a Cauchy data set for which a solution with $M>0$
exists, then there is no solution with $M>0$ for any
other Cauchy data set within the family.
\end{proposition}

\section{Conclusions}
The combination of propositions \ref{res:dirichlet} and \ref{res:fix}
with the boundary data that a given interior infers
onto the exterior problem, as shown in Section \ref{sec:bc},
leads us to:
\begin{theorem}
Let $\spI$ be a given stationary and axially symmetric region
of spacetime containing an electromagnetic field,
and bounded by a symmetry-preserving and simply connected
hypersurface $\sup$, spatially homeomorphic to a two-sphere,
and containing no surface charge currents.
If $\spI$ can be matched across $\sup$
to an asymptotically flat (with $M> 0$)
stationary and axially symmetric electrovacuum 
Einstein-Maxwell field 
$(\mm2^\E,g^\E,\bm\ff)$, preserving the symmetry,
and where the identification on $\sup$ has been prescribed, 
then $(\mm2^\E,g^\E,\bm\ff)$ is unique within the set of positive
mass solutions.\fiproofns
\end{theorem}
As a first remark, the same result holds if $\sup$ is allowed to
contain surface
electromagnetic charge currents, as long
as these currents are taken as given data.
Secondly, although for physically motivated situations the restriction
of the uniqueness result to the set of
positive mass solutions is irrelevant,
it may be desirable to have a complete uniqueness result.
This might be expected to follow by using existence considerations,
and it is currently under study.

All in all, this final result can be stated in other words as follows:
under natural assumptions on $\sup$
--preservation of the symmetries and the prescription of the identification
of both sides of $\sup$ \cite{MASEuni}--,
the exterior electrovacuum
field generated by an isolated distribution of charged matter
in axial rotation and in equilibrium is unique, provided that it exists.

Some comments on the definition of asymptotic flatness are in order here.
The asymptotic flatness has \emph{explicitly} entered the problem
only through the use of the asymptotic conditions
(\ref{eq:uomegainfinity})-(\ref{eq:potinfinity}).
In principle, one could have included NUT and monopole
charges in the expressions for $U$ and $\Lambda$ at infinity.
In fact, it can be shown that the inclusion of NUT and monopole
charges does not change the present results, provided we can use
Weyl coordinates globally on $\spE$ (see (\ref{eq:ds2e1})).
Nevertheless, the existence of global Weyl coordinates
has been only ensured when $\spE$ admits an asymptotically flat
four-end \cite{rho} (see e.g. \cite{Chrusciel94} for definitions).
Therefore, this is the definition of asymptotic flatness
that has been implicitly used all throughout the paper,
which implies the vanishing of the NUT and monopole
charges \cite{walter84},
as mentioned in Section \ref{sec:ext}.

Looking at the procedure used here, several generalisations 
to include different fields on the exterior region could be
attempted quite straightforwardly.
The starting point would be the possibility of using the
Weyl coordinates (orthogonal transitivity and a trace-free energy-momentum
tensor \cite{sol}). Then, one should consider the boundary conditions
for the fields on the matching hypersurface, to be
combined to those inferred by the spacetime matching conditions,
according to Proposition \ref{res:matching},
in order to find the form of the 
Cauchy boundary data for the exterior problem.
Of course, the main ingredient would consist of picking up fields
that can be described by $\sigma$-models \cite{smodels},
and use then the Mazur construction for the Dirichlet problem,
and the divergence-free currents for the fixing of the Dirichlet
data with the whole set of Cauchy data.

\ack
 I am grateful to Marc Mars, Jos\'e Senovilla and Brien Nolan for
 those many discussions, comments and ideas, and
 for the careful readings of the manuscript.
 This work was mainly produced at the School of Mathematical Sciences,
 Queen Mary, University of London, funded by the EPSRC, Ref. GR/R53685/01.
 I thank the the Irish Research Council for Science, Engineering
 and Technology (IRCSET), for
 postdoctoral fellowship Ref. PD/2002/108.

\appendix

\section{The explicit matching}
\label{sec:matching}
This section is devoted to explicitly present the general
matching of two stationary and axially symmetric spacetimes (with boundary),
$\spIB$ and $\spEB$,
across a matching hypersurface $\sup~(\equiv\sup^\I=\sup^\E)$
preserving the stationarity and the axial symmetry.
The latter implies that $\sup$ is timelike.
Furthermore, we impose on $\spE$ the following two requirements:
(i) the $G_2$ on $T_2$ group of isometries acts orthogonally transitively (OT),
and (ii) the trace of the Einstein tensor is zero.
These two conditions
simply account for having the possibility of simplifying the problem
by 
using the Weyl coordinates (at least locally) (see \cite{sol}).

Apart from preserving the symmetries, the only implicit
assumption we are making on
$\sup$ is that it intersects points at the axis,
so that the axis intersects both $\mm2^\I$ and $\mm2^\E$,
which is ensured if $\sup$ is spatially homeomorphic to a two-sphere.
Conditions (i) and (ii) are satisfied when $\spE$ contains
a stationary and axisymmetric electrovacuum field \cite{sol}.



The equations for the matching
with a vacuum $\spE$ spacetime where presented in an explicit manner in
\cite{conv}. The reader is referred to that work
(and also to \cite{MASEuni,merce} for the OT $\spI$ case) for further details.

There exists a coordinate system $\{T,\Phi,r,\zeta\}$ in which
the line-element for $g^\I$ reads \cite{WW1,sol}
\begin{equation}
\fl \rmd s^2_\I=-e^{2V}\left(\rmd T+ B \rmd \Phi +
\ww ~ \rmd\zeta\right)^2+e^{-2V}
\left[e^{2h}\left(\rmd r^2+\rmd \zeta^2\right)+
\alpha^2\left(\rmd\Phi+\www ~ \rmd\zeta\right)^2\right],
\label{eq:ds2i}
\end{equation}
where $V$, $B$, $h$, $\ww$, $\www$ and $\alpha$ are functions
of $r$ and $\zeta$, and $\vec\eta^{\,\I}=\partial_\Phi$ is the
(intrinsically defined) axial Killing vector.
The vector ${\vec \xi}^{\,\I} \equiv\partial_T$ is a stationary Killing vector.
By assumption, $\spE$ admits Weyl coordinates $\{t,\phi,\rho,z\}$,
in which its line-element is given by (\ref{eq:ds2e1}),
the intrinsically defined
axial Killing vector is given by $\vec\eta^{\,\E}=\partial_\phi$,
and we take $\vec\xi^{\,\E}=\partial_t$.
As mentioned in Section \ref{sec:ext}
it is straightforward to show that
the most general matching hypersurface preserving stationarity and
axial symmetry $\sup$,
as seen form the $\spE$ side, is given in parametric form as
\cite{MASEuni,mps,rho}
\[
\sup^\E:\{ t=\tau,\phi=\varphi,\rho=\rho(\mu),z=z(\mu)\},
\]
where $\{\tau,\varphi,\mu\}$ parametrise $\sup$. 
At this point, the matching procedure
introduces two degrees of freedom, a product of the identification
of $\partial_t$ with any linear combination
$a (\partial_T+ b\partial_\Phi)$ on $\sup$ \cite{MASEuni,mps,conv}.
Of course, this is of relevance if $\partial_t$ is intrinsically defined,
as it happens, for instance, for asymptotically flat $\spE$ \cite{MASEuni}.
Despite that, one can proceed with the matching avoiding the
explicit appearance of $a,b$ by absorbing them in $\spI$,
changing coordinates in $\spI$ to a new set of ``primed''
coordinates, with corresponding functions
$V'$, $B'$, $h'$, $\ww'$, $\www'$ and $\alpha'$ in (\ref{eq:ds2i}),
so that $\partial_{T'}=a (\partial_T+ b\partial_\Phi)$
while keeping $\partial_{\Phi'}=\partial_{\Phi}$
(see \cite{conv} for the explicit relations
between primed and the original non-primed quantities).
Nevertheless, if the interior geometry (\ref{eq:ds2i})
was given beforehand, these relations involving the parameters
$a,b$ have to be taken into account.
Having this in mind, and dropping the primes,
the most general parametric
form of $\sup$ as seen from $\spI$ reads then \cite{conv}
\[
  \sup^\I:\{T=\tau+f_T(\mu),
\Phi=\varphi+f_\Phi(\mu),r=r(\mu),\zeta=\zeta(\mu)\},
\]
where $\dot f_T(\mu)=\dot \zeta (B \www -\ww)(\mu)$ and
$\dot f_\Phi(\mu)=-\dot \zeta \www(\mu)$.

Two basis for the tangent spaces to $\sup$,
from the $\sup^\I$ and $\sup^\E$ sides, $\{\vec e^{\,\I}_a\}$ and
$\{\vec e^{\,\E}_a\}$ $a:1,2,3$ respectively, which
are to be identified eventually as $\{\vec e_a\}$
in the final matched spacetime
\cite{MASEhyper}, read explicitly
\begin{eqnarray*}
\vec e^{\,\I}_1=\vec \xi^{\,\I}|_{\sup},~~~
\vec e^{\,\I}_2=\vec \eta^{\,\I}|_{\sup},~~~
\vec e^{\,\I}_3=\dot r \partial_r+\dot\zeta\left[(B \www -\ww)\partial_T-
\www \partial_\Phi + \partial_\zeta\right]|_\sup,\\
\vec e^{\,\E}_1=\vec \xi^{\,\E}|_{\sup},~~~
\vec e^{\,\E}_2=\vec \eta^{\,\E}|_{\sup},~~~
\vec e^{\,\E}_3=\dot \rho\,\partial_\rho+\dot z\,\partial_z|_\sup.
\end{eqnarray*}
In principle, any choice of normal vector $\bm n$ is suitable for computing
the matching conditions, but it is fixed, by convention, so that it points
$\spIB$ inwards and $\spEB$ outwards, and by convenience, that
its norm is the same as that of $\vec e_3$.
On the $\sup^\E$ side it reads
$\bm n^\E=e^{2(k-U)}(-\dot z\,\d\rho+\dot\rho\,\d z)$,
while on the $\sup^\I$ side, this is
$\bm n^\I=\epsilon\, e^{2(h-V)} (-\dot \zeta\,\d r+\dot r\,\d\zeta)$,
where $\epsilon$ is a sign.
The equality of the first fundamental forms is equivalent to
the following four equations, the so-called
preliminary junction conditions,
\begin{eqnarray}
  \label{eq:pjc1}
  U|_\sup=V|_\sup, ~~ A|_\sup=B|_\sup,\\
  \label{eq:pjc2}
  \rho|_\sup=\alpha|_\sup,\\
  \label{eq:pjc3}
  e^{2k}|_\sup(\dot\rho^2+\dot z^2)=e^{2h}|_\sup(\dot r^2+\dot \zeta^2).
\end{eqnarray}
Since $\sup$ is timelike
the remaining matching conditions simply consist
of equating the second fundamental forms,
$K_{ab}\equiv e_a{}^\alpha e_b{}^\beta\nabla_\alpha n_\beta$,
which are symmetric \cite{MASEhyper}, as computed from $\sup^\E$
and $\sup^\I$.
Labelling by $A,B=1,2$ the components corresponding to the
orbits, 
the set of six 
equations $K^\E_{ab}=K^\I_{ab}$,
after using (\ref{eq:pjc1})-(\ref{eq:pjc3}), can be expressed and
grouped as follows:
\begin{eqnarray}
  \label{eq:KAB_1}
  \fl K^\E_{AB}=K^\I_{AB}:\hspace{1cm}&&\vec n^\E(U)|_\sup=\vec n^\I(V)|_\sup,
  ~~\vec n^\E(A)|_\sup=\vec n^\I(B)|_\sup,\\
  \label{eq:KAB_2}
  &&\dot z=-\vec n^\I(\alpha)|_\sup\\
  \label{eq:KA3}
  \fl K^\E_{A3}=K^\I_{A3}:&& \partial_r(B\www-\ww)|_\sup=0,~~
  \partial_r\www|_\sup=0\\
  \fl K^\E_{33}=K^\I_{33}:&&
  e^{2k}(\dot\rho^2+\dot z^2)\vec n^{\E}(k)+e^{2(k-U)}
  \left(\ddot\rho\dot z-\dot\rho\ddot z\right)|_{\sup}\nonumber\\
  \label{eq:K33}
 && ~~~=e^{2h}(\dot r^2+\dot \zeta^2)\vec n^{\I}(h)+e^{2(h-V)}
  \left(\ddot r\dot \zeta-\dot r \ddot \zeta\right)|_{\sup},
\end{eqnarray}
where (\ref{eq:KAB_1}) has been used in (\ref{eq:K33}).

Using all the equations one finds first that
$\dot\rho^2+\dot z^2=(\alpha_{,r}^2+\alpha_{,\zeta}^2)|_\sup
(\dot r^2+\dot\zeta^2)$, meaning that
$\alpha_{,r}^2+\alpha_{,\zeta}^2$ cannot vanish anywhere
on $\sup$, and secondly, that
(\ref{eq:pjc3}) and (\ref{eq:K33}) can be substituted by
\begin{eqnarray}
  \label{eq:pjc3bis}
  k|_\sup=\left.h-\frac{1}{2}
    \log(\alpha_{,r}^2+\alpha_{,\zeta}^2)\right|_\sup, \\
  \label{eq:K33bis}
  \vec n^\E(k)|_\sup=\left.\vec n^\I(h)-\epsilon\,
  \frac{\alpha_{,r}\dot\alpha_{,\zeta}-\alpha_{,\zeta}\dot\alpha_{,r}}
  {\alpha_{,r}^2+\alpha_{,\zeta}^2}\right|_\sup,
\end{eqnarray}
respectively. A straightforward calculation shows that,
once the rest of the equations hold,
equations (\ref{eq:K33bis}) and the derivative of (\ref{eq:pjc3bis})
along $\mu$ are equivalent to 
\begin{equation}
    \label{eq:isra1}
    n^{\E\,\alpha} n^{\E\,\beta} S^{\E}_{\alpha\beta}|_\sup=
    n^{\I\,\alpha} n^{\I\,\beta} S^{\I}_{\alpha\beta}|_\sup,~~~
    n^{\E\,\alpha} e_3^{\E\,\beta} S^{\E}_{\alpha\beta}|_\sup=
    n^{\I\,\alpha} e_3^{\I\,\beta} S^{\I}_{\alpha\beta}|_\sup,
\end{equation}
where $S_{\alpha\beta}$ denotes the Einstein tensor.
On the other hand, the equations in (\ref{eq:KA3})
are equivalent to \cite{conv}
\begin{equation*}
*(\bm \xi^\I\wedge \bm\eta^\I\wedge \d \bm \xi^\I)|_\sup=0=
*(\bm \xi^\I\wedge \bm\eta^\I\wedge \d \bm \eta^\I)|_\sup,
\end{equation*}
(see also \cite{mps}), and, in turn, 
equivalent to \cite{conv} 
(up to additive contants, but these vanish because $\sup$
intersects the axis) 
\begin{equation}
    \label{eq:isra2}
  n^{\I\,\alpha} e_1^{\I\,\beta} S^{\I}_{\alpha\beta}|_\sup=0,~~~
  n^{\I\,\alpha} e_2^{\I\,\beta} S^{\I}_{\alpha\beta}|_\sup=0.
\end{equation}
Since $n^{\I\,\alpha} e_1^{\E\,\beta} S^{\E}_{\alpha\beta}=0=
n^{\I\,\alpha} e_2^{\E\,\beta} S^{\E}_{\alpha\beta}$
are identically satisfied due to the OT structure of the
$\spE$ region,
the two equations in (\ref{eq:isra2}) together with
those in (\ref{eq:isra1}) constitute the
set of the four Israel conditions.

To sum up, the whole set of matching conditions is formed by
the ten equations in (\ref{eq:pjc1}), (\ref{eq:pjc2}),
(\ref{eq:KAB_1}), (\ref{eq:KAB_2}),
(\ref{eq:pjc3bis}), (\ref{eq:K33bis}),
and (\ref{eq:isra2}), taking into account that
up to an additive constant in $k|_\sup$, equations 
(\ref{eq:pjc3bis})-(\ref{eq:K33bis}) can be substituted by
(\ref{eq:isra1}).
If required, that constant should then be fixed
by considering (\ref{eq:pjc3bis}).

In principle, if no conditions are imposed on the matter content
in $\spE$, equations
(\ref{eq:pjc3bis})-(\ref{eq:K33bis}),
or equivalently 
(\ref{eq:isra1}), 
provide conditions
on $\sup$ for the function $k$, given the interior geometry.
Conversely, if one imposes conditions on the matter content
in $\spE$, i.e. on $S^{\E}_{\alpha\beta}$,
so that the function $k$
has to satisfy equations involving the rest of the functions,
equations 
(\ref{eq:pjc3bis})-(\ref{eq:K33bis})
may represent conditions on the $\spI$ side only.
In particular, this becomes clear when $\spE$ is vacuum,
because (\ref{eq:isra1}) only involve quantities in $\spI$,
and therefore (\ref{eq:pjc3bis})-(\ref{eq:K33bis}) consist, in fact,
of conditions on $\sup^\I$ (apart from providing the value of the
additional constant in $k$).

Although not as explicitly apparent as in the vacuum case,
it is straightforward to show that the same holds true
when $\spE$ is an electrovacuum solution
whose electromagnetic field extends continuously across $\sup$
onto $\spI$.
As mentioned in Section \ref{sec:ext},
$k$ is determined, up to an additive constant, by quadratures
in terms of the rest of the metric functions, which satisfy the Ernst
equations.
Defining $\bm \ernstf\equiv -\d N + i \bm w$, where this and
the following quantities are referred to $\kk=\vec\xi$, 
the quadratures for $k$ in (\ref{eq:ds2e1})
can be cast as (see \cite{sol,heusler})
\begin{eqnarray*}
   \frac{1}{\rho}k_{,\rho}= \frac{1}{4 N^2}
   \left(\ernstf_\rho\bar \ernstf_\rho-\ernstf_z\bar \ernstf_z\right)+
   \frac{1}{N}\left(\Lambda_{,\rho}\bar\Lambda_{,\rho}-
     \Lambda_{,z}\bar\Lambda_{,z}\right),\\
   \frac{1}{\rho}k_{,z}= \frac{1}{4 N^2}
   \left(\ernstf_\rho\bar \ernstf_z+\ernstf_z\bar \ernstf_\rho\right)+
   \frac{1}{N}\left(\Lambda_{,\rho}\bar\Lambda_{,z}+
     \Lambda_{,z}\bar\Lambda_{,\rho}\right).
\end{eqnarray*}
The explicit expressions in terms of $U$ and $A$ are
then obtained by using (\ref{eq:UN})-(\ref{eq:Aomega}),
and
a straightforward calculation on $\sup$ leads to
\begin{eqnarray*}
  \dot k|_{\sup}=\left.\frac{\rho}{\dot\rho^2+\dot z^2}
  \left(\aa\dot\rho-\bb\dot z\right)\right|_{\sup},~~~
  \vec n^\E(k)|_{\sup}=\left.\frac{\rho}{\dot\rho^2+\dot z^2}
    \left(\aa\dot z+\bb\dot \rho\right)\right|_{\sup},
\end{eqnarray*}
where we have defined
\begin{eqnarray*}
  \aa\equiv\left.\dot U^2+\vec n^\E(U)^2+\frac{e^{4U}}{4\rho^2}
  \left(\dot A^2+\vec n^\E(A)^2\right)-e^{-2U}
  \left(\dot\Lambda\dot{\bar\Lambda}-\vec n^\E(\Lambda)\vec n^\E(\bar\Lambda)
  \right)\right|_{\sup},\\
  \bb\equiv\left.2\dot U\vec n^\E(U)-\frac{e^{4U}}{2\rho^2}\dot A\vec n^\E(A)-
    e^{-2U}\left(\vec n^\E(\Lambda)\dot{\bar\Lambda}+\dot\Lambda\vec n^\E(\bar\Lambda)\right)
    \right|_{\sup}.
\end{eqnarray*}
The continuity of the electromagnetic field across $\sup$
implies relations (\ref{eq:dlambda_exp})
for $\dot\Lambda|_\sup$ and
$\vec n^\E( \Lambda)|_\sup$ in terms of (I) quantities.
Using these relations together with
(\ref{eq:pjc1}), (\ref{eq:pjc2}), (\ref{eq:KAB_1}) and
(\ref{eq:KAB_2}),
$\dot k|_\sup$ and $\vec n^\E(k)|_\sup$ 
can be obtained in terms of quantities of $\spI$,
and hence obtain expressions for
(\ref{eq:K33bis}) and the derivative of (\ref{eq:pjc3bis})
along $\mu$ as conditions for quantities on the $\spI$ side only.
Equivalently, equations (\ref{eq:isra1}), within the set of
matching conditions, provide conditions on $\spI$ only.

On the other hand, of course, it must be stressed that
conditions on the matter content in $\spE$
only translate to conditions on the $\spI$ geometry
through the four Israel conditions (\ref{eq:isra1})-(\ref{eq:isra2}),
and thus, given an interior geometry satisfying these conditions,
the rest of conditions involve the form of the matching
hypersurface at the $\E$ side, $\sup^\E$, plus boundary data (of Cauchy type)
for $U$ and $A$ on $\sup^\E$ given the interior geometry.

All the above can be summarised as follows:
\begin{proposition}
\label{res:matching}
Let $\spE$ and $\spI$ be two stationary and axially symmetric
spacetimes which are to be matched across their respective
symmetry-preserving boundaries $\sup^\E$ and $\sup^\I$,
whose identification is denoted as $\sup (\equiv\sup^\E=\sup^\I)$.
Let us also assume that $\spE$ contains a stationary and axially
symmetric electromagnetic field and satisfies the Einstein-Maxwell
equations without sources.
The whole set of matching conditions can be reorganised as follows:
\begin{enumerate}
\item[(i)] Conditions on $\sup^\I$, given by
(\ref{eq:isra1}), (\ref{eq:isra2}),
which form, in principle, an overdetermined system for $\{r(\mu),\zeta(\mu)\}$.
If a solution exists the matching is possible (locally) and the hypersurface
$\sup^\I$ is determined generically.
\item[(ii)] Equations defining $\sup^\E$, given by (\ref{eq:pjc2}),
(\ref{eq:KAB_2}).
\item[(iii)] Boundary conditions for the electrovacuum problem in $\spE$,
given by (\ref{eq:pjc1}), (\ref{eq:KAB_1}). If required,
the additive constant in $k$ is determined by using
(\ref{eq:pjc3bis}).\fiproofns
\end{enumerate} 
\end{proposition}
\begin{remarkpro}
The explicit expressions apply for the definitions and metric functions
as introduced above in this Appendix, but can be easily reexpressed
in an intrinsic manner.
\end{remarkpro}

\section{The divergence-free forms}
\label{app:jj}
The explicit expressions of the divergence free (in $(\mm2,g)$)
1-forms used for the decomposition of the su$(2,1)$-valued 1-form
$\bm\jj_{\kk}$ (for either $\kk$)
are given as follows, both in terms of $\ernst$ and $\Lambda$
and in terms of the sometimes more convenient $N$,
$\Lambda$, $\Omega$ and
$\bm w$ (for $\d \Omega$ (\ref{eq:domega})):
\begin{eqnarray*}
&&\fl  \bm\jj^e=-\frac{i}{2N^2}\left[2
(\bar\Lambda\d\Lambda- \Lambda\d\bar\Lambda)-\d(\bar\ernst-\ernst)\right]
=\frac{1}{N^2}\bm w,\\
&&\fl  \bm\jj^d=\frac{1}{2N^2}\left[\d(\ernst\bar\ernst)
+\Lambda\bar\Lambda\d(\ernst+\bar\ernst)-
(\bar\Lambda\d\Lambda- \Lambda\d\bar\Lambda)(\bar\ernst-\ernst)\right]\\
&&~~~~
\fl=\frac{1}{N^2}\left[\Omega\bm w+N(\d N +\d(\Lambda\bar\Lambda))\right],\\
&&\fl  \bm\jj^s=-\frac{3i}{4N^2}\left[
\Lambda\bar\Lambda\d(\ernst-\bar\ernst)-
(\bar\Lambda\d\Lambda- \Lambda\d\bar\Lambda)(\ernst+\bar\ernst)\right]\\
&&\fl~~~~
=\frac{3}{2N^2}\left[\Lambda\bar\Lambda\bm w-
i N(\bar\Lambda\d\Lambda- \Lambda\d\bar\Lambda)\right],\\
&&\fl  \bm\jj^p=\frac{i}{2N^2}\left[2\ernst\bar\ernst
(\bar\Lambda\d\Lambda- \Lambda\d\bar\Lambda)
+(\ernst+2\Lambda\bar\Lambda)\ernst\d\bar\ernst
-(\bar\ernst+2\Lambda\bar\Lambda)\bar\ernst\d\ernst\right]\\
&&\fl~~~~
=\frac{1}{N^2}\left\{
(N^2-\Omega^2-\Lambda^2\bar\Lambda^2)\bm w
-2N\Omega\,\d N-4\Rea\left[\left(\Omega-i(N+\Lambda\bar\Lambda)\right)
N \bar\Lambda\d\Lambda\right]\right\},\\
&&\fl   \bm\jj^h=\frac{1}{2N^2}\left[
\bar\Lambda\d(\ernst-\bar\ernst)+(\ernst+\bar\ernst)\d \bar\Lambda+
2\bar\Lambda^2\d \Lambda\right]=\frac{1}{N^2}
\left[i\bar\Lambda\bm w-N\d \bar\Lambda\right],\\
&&\fl  \bm\jj^a=\frac{1}{2 N^2}\left[
\ernst(\ernst+\bar\ernst)\d \bar\Lambda+2\bar\Lambda^2
(\ernst\d \Lambda-\Lambda\d \ernst)-\bar\Lambda\d(\ernst\bar\ernst)\right]\\
&&\fl~~~~
=\frac{1}{N^2}\left\{N^2\d\bar\Lambda-N\bar\Lambda\d N
-2N\bar\Lambda^2\d\Lambda-
(\Omega+i\Lambda\bar\Lambda)(\bar\Lambda\bm w+iN\d\bar\Lambda)
\right\}.
\end{eqnarray*}
The vacuum case is recovered by $\Lambda=0$ (so that $\d \Omega=\bm w$),
and thus
only three 1-forms survive, namely $\bm\jj^e$, $\bm\jj^d$ and
$\bm\jj^k$. Their expressions in the case $\kk=\vec\xi$,
and after the substitution $N=-e^{2U}$,
read 
$\bm\jj^e=e^{-4U}\d\Omega$,
$\bm\jj^d=2\d U+e^{-4U}\Omega\d\Omega$, and
$\bm\jj^p=-4\Omega\d U+(1-e^{-4U}\Omega^2)\d\Omega$,
which were those used in constructing the divergence-free
1-form in \cite{MASEuni}.

With the above expressions one is ready to compute the relations
(\ref{eq:relations}) by using the differences in the Cauchy data
of the two solutions in terms of
$\Lambda$, $N$, $\Omega$ and $\bm w$ (\ref{eq:mcem})-(\ref{eq:mctwist}):
\begin{eqnarray*}
&\Lambda_{(2)}|_{\curv_\sup}=\Lambda_{(1)}|_{\curv_\sup}+\lambda,
&\d \Lambda_{(2)}|_{\curv_\sup}=\d \Lambda_{(1)}|_{\curv_\sup},\\
&N_{(2)}|_{\curv_\sup}=N_{(1)}|_{\curv_\sup},
&\d N_{(2)}|_{\curv_\sup}=\d N_{(1)}|_{\curv_\sup},\\
&\Omega_{(2)}|_{{\curv_\sup}}=
\Omega_{(1)}|_{{\curv_\sup}}+\comega +i(\bar\lambda \Lambda_{(1)}-
   \lambda\bar\Lambda_{(1)})|_{\curv_\sup},~~~
&\bm w_{(2)}|_{\curv_\sup}=\bm w_{(1)}|_{\curv_\sup}.
\end{eqnarray*}
Indeed, (\ref{eq:ws}) is recovered with
\begin{eqnarray*}
\fl  \hat c_p=c_p,~~
  \hat c_{a1}=c_{a1}+i2c_p\lambda ,~~
  \hat c_{a2}=c_{a2}-i2 c_p\bar\lambda \\
\fl  \hat c_d=c_d+ c_{a1}\bar\lambda+c_{a2}\lambda+2c_p\,\comega,~~
  \hat c_s=c_s+3i(c_{a1}\bar\lambda-c_{a2}\lambda)-
  6\,c_p\,\lambda\bar\lambda,\\
\fl  \hat c_{h_1}=c_{h_1}+(c_d+ic_s)\lambda-c_{a_1}(\lambda\bar\lambda
+\comega i)+2\lambda^2(c_{a_2}-ic_p\bar\lambda)+2c_p\lambda \comega,\\
\fl  \hat c_{h_2}=c_{h_2}+(c_d-ic_s)\bar\lambda-c_{a_2}(\lambda\bar\lambda
-\comega i)+2\bar\lambda^2(c_{a_1}+ic_p\lambda)+2c_p\bar\lambda \comega,\\
\fl  \hat c_e=c_e+i(c_{h_2}\lambda-c_{h_1}\bar\lambda)
+\lambda\bar\lambda(ic_{a_1}\bar\lambda-ic_{a_2}\lambda-
c_p \lambda \bar\lambda+c_s)
-(c_d+c_{a_1}\bar\lambda+c_{a_2}\lambda+c_p \comega)\comega.
\end{eqnarray*}
These expressions are then introduced into (\ref{eq:final}),
which has to hold for arbitrary $\{c_e,\ldots,c_s\}$.
Putting $c_{h_1}=1$ and the rest zero (say, the $c_{h_1}$-equation:
in what follows, the analogous procedure will be referred to as
$c_{x}$-equations) 
one immediately obtains
\[
Q_{(1)}=Q_{(2)},
\]
whereas the equation involving $c_s$ implies
\[
Q_{(2)}(\bar\lambda -\lambda)=0.
\]
Using these two relations on a suitable combination of the
equations involving $c_{a_1}$ and $c_d$, explicitly
$\lambda$[$\bar\lambda$($c_d$-equation)$-2$ ($c_{a_1}$-equation)],
one obtains
\[
2\lambda\bar\lambda(M_{(1)}+M_{(2)})=Q_{(2)}
\comega i (\lambda+\bar\lambda),
\]
whose real part reads simply
\[
2\lambda\bar\lambda(M_{(1)}+M_{(2)})=0.
\]
The same procedure applied to another combination of the equations
involving  $c_{d}$ and $c_{p}$, explicitly
$\comega$($c_d$-equation)$-2$ ($c_{p}$-equation),
leads to
\[
2 \comega (M_{(1)}+M_{(2)})=-\comega\, Q_{(2)}
(\lambda+\bar\lambda).
\]
From the last two equations,
if $M_{(1)}+M_{(2)}\neq 0$, then $\lambda=\comega=0$,
and thus the main result in Section \ref{sec:fixing} follows.
As expected, once $\lambda=\comega=0$ it is immediate to see that then
$M_{(1)}=M_{(2)}$.

It must be stressed here that the case $M_{(1)}+M_{(2)}=0$ does not
lead to the same result: by setting $M_{(2)}=-M_{(1)}\neq 0$, it is
immediate to find that $\comega=0$, $\lambda-\bar\lambda=0$
(so that only the real part of $\lambda$ survives) and finally,
$\lambda=2M_{(1)}/Q_{(1)}$ exhausting equation (\ref{eq:final}).


\section*{References}


\begin{thebibliography}{99}


\bibitem{noconv2}Ioka K and Sasaki M \Journal{\PRD}{67}
{124026}{}{2003}
{Grad-Shafranov equation in noncircular stationary axisymmetric spacetimes}

\bibitem{MASEuni}Mars M and Senovilla J M M \Journal{\MPL}{A13}
{1509}{-1519}{1998}
{On the construction of global models describing
rotating bodies; uniqueness of the exterior gravitational field}

\bibitem{Ori}Ori A \Journal{\CQG}{7}
{985}{-998}{1990}
{The general solution for spherical charged dust}

\bibitem{evsph} Fayos F, Senovilla J M M and Torres R \Journal{\CQG}{20}
{2579}{-2594}{2003}
{Spherically symmetric models for charged stars and voids: I Charge bound}

\bibitem{sol}Stephani H, Kramer D, MacCallum M A H,
Hoenselaers C and Herlt E (2003)
{\em Exact solutions of Einstein's field equations. Second Edition}, Cambridge
University Press, Cambridge

\bibitem{heusler}Heusler M (1996) {\it Black Hole Uniqueness Theorems\/},
Cambridge lecture notes in Physics, Cambridge Univ. Press, Cambridge

\bibitem{conv}Vera R \Journal{\CQG}{20}
{2785}{-2791}{2003}
{Influence of general convective motions on the exterior
of isolated rotating bodies in equilibrium} 

\bibitem{jackson} Jackson J D \textit{Classical Electrodynamics}
(Second Edition) John Wiley \& Sons Inc (1975)

\bibitem{maseaxconf}Mars M and Senovilla J M M \Journal{\CQG}{10}
{1633}{-1647}{1993}
{Axial symmetry and conformal Killing vectors}

\bibitem{carter70}Carter B \Journal{\CMP}{17}
{233}{-238}{1970}
{The commutation property of a stationary axisymmetric system}

\bibitem{mps}Vera R \Journal{\CQG}{19}
{5249}{-5264}{2002}
{Symmetry-preserving matchings}

\bibitem{carterbh}Carter B \Journal{\CMP}{99}
{563}{-591}{1985}
{Bunting identity and Mazur identity for non-linear elliptic
systems including the black hole equilibrium problem}

\bibitem{weinstein1}Weinstein G \Journal{\CPAM}{43}
{903}{-948}{1990}
{On rotating black holes in equilibrium in General Relativity}

\bibitem{rho} Vera R In preparation

\bibitem{harrison} Harrison B K \Journal{\JMP}{9}
{1744}{-1752}{1968}
{New solutions of the Einstein-Maxwell equations from old}

\bibitem{ernst68b} Ernst F J \Journal{\PR}{168}
{1415}{-1417}{1968}
{New formulation of the axially symmetric gravitational field problem II}

\bibitem{walter84} Simon W \Journal{\JMP}{25}
{1035}{-1038}{1984}
{The multipole expansion of stationary Einstein-Maxwell fields}

\bibitem{bmg}Breitenlohner P, Maison D and Gibbons G \Journal{\CMP}{120}
{295}{-333}{1988}
{4-Dimensional Black Holes from Kaluza-Klein theories}

\bibitem{smodels}Breitenlohner P and Maison D \Journal{\CMP}{209}
{785}{-810}{2000}
{On nonlinear $\sigma$-models arising in (super)-gravity}

\bibitem{carterww} Carter B ``Black hole equilibrium states'' in
{\it Black Holes}, Ed. B. De Witt, C. De Witt, (NewYork: Gordon and Breach)
1973

\bibitem{Chrusciel94}Chrusciel P T `` 'No hair' theorems - folklore,
conjectures, results'' in {\it Proceedings of the joint AMS/CMS
Conference on mathematical physics and differential geometry,
August 1993, Vancouver} eds. J. Beem and K.L. Duggal,
{\it Cont. Math.} (1994), 23-49 (gr-qc/9402032)

\bibitem{merce}Mart\'{\i}n-Prats M M and Senovilla J M M 1993 in
{\it Rotating Objects and Relativistic Physics}, ed. F.J. Chinea and
L.M. Gonz\'alez-Romero, 
{\it Lectures Notes in Physics} {\bf 423}, (Berlin: Springer-Verlag) p 136;
see also Mart\'{\i}n-Prats M M, Ph. D. Thesis, Universitat de Barcelona, 1995

\bibitem{WW1}Wainwright J \Journal{\JPA}{14}
{1131}{-1147}{1981}
{Exact spatially inhomogeneous cosmologies}


\bibitem{MASEhyper}Mars M and Senovilla J M M \Journal{\CQG}{10}
{1865}{-1897}{1993}
{Geometry of general hypersurfaces in spacetime: junction conditions}




\end{thebibliography}
\end{document}